\documentclass{aa} 
\usepackage{rotating}
\usepackage[colorlinks=true, citecolor=blue, urlcolor=blue]{hyperref}

\usepackage{graphicx}
\usepackage{txfonts}
\usepackage{amssymb}
\usepackage{lscape}
\usepackage[bf,nooneline]{subfigure}
\bibpunct{(}{)}{;}{a}{}{,}

\begin{document} 

\title{Lower mass normalization of the stellar initial mass function for
dense massive early-type galaxies at z $\sim$ 1.4}

   \author{A. Gargiulo\inst{1}\thanks{adriana.gargiulo@brera.inaf.it}, P. Saracco\inst{1}, M. Longhetti\inst{1}, S. Tamburri\inst{1,2}, I. Lonoce\inst{1,2}, 
F. Ciocca\inst{1,2} }

   \institute{INAF - Osservatorio Astronomico di Brera, Via Brera 28, 20121 Milan, Italy\\
         \and Dipartimento di Scienza e Alta Tecnologia, Universit\`{a} degli Studi dell'Insubria, via Valleggio 11, 22100 Como, Italy}

   \date{Received 20 May 2014 / Accepted 16 October 2014 }

 
  \abstract
{}
   {This paper aims at understanding whether the normalization of the stellar initial mass function (IMF) of massive 
galaxies varies with cosmic time and/or with mean stellar mass density $\Sigma$ = M$_{\star}$/(2$\pi$R$_{e}^{2}$).}
{We have tackled this question by taking advantage of a spectroscopic sample of 18 dense ($\Sigma > $2500M$_{\odot}$pc$^{-2}$) 
massive early-type galaxies (ETGs) that we collected at 1.2 $\lesssim z \lesssim$ 1.6. 
Each galaxy in the sample was selected in order to have available: i) a high-resolution deep HST-F160W image to 
visually classify it as an ETG; ii) an accurate velocity dispersion estimate; iii) stellar mass derived through the fit of 
multiband photometry; and iv) structural parameters (i.e. effective radius R$_{e}$ and Sersic index n) derived in the F160W-band. 
We have constrained the mass-normalization of the IMF of dense high-z ETGs by comparing the true stellar masses of the ETGs 
in the sample (M$_{true}$) derived through virial theorem, hence IMF independent, with those inferred through
the fit of the photometry which assume a reference IMF (M$_{ref}$). Adopting the virial estimator as proxy of the true 
stellar mass, we have implicitly assumed that these systems have zero dark matter. However, 
recent dynamical analysis of massive local ETGs have shown that the dark matter fraction within R$_{e}$
in dense ETGs is negligible (<5-10$\%$) and simulations of dissipationless mergers of spheroidal galaxies 
have shown that this fraction decreases going back with time.
Accurate dynamical models of local ETGs performed by the ATLAS$^{3D}$ team 
have shown that the virial estimator is prone to underestimating or overestimating the total masses. We have considered this, 
and based on the results of ATLAS$^{3D}$ we have shown that for dense ETGs the mean value of total masses derived 
through the virial estimator with a non-homologous virial coefficient and Sersic-R$_{e}$ 
are perfectly in agreement with the mean value of those derived through more sophisticated dynamical models, 
although, of course, the estimates show higher uncertainties.}
{Tracing the variation of the parameter $\Gamma$ = M$_{true}$/M$_{ref}$ with velocity dispersion $\sigma_{e}$, 
we have found that, on average, dense ETGs at $<z>$ = 1.4 follow the same IMF-$\sigma_{e}$ trend 
of typical local ETGs, but with a lower mass-normalization. 
The observed lower normalization 
could be evidence of $i)$ an evolution
of the IMF with time or $ii)$ a correlation with $\Sigma$.
To discriminate between the two possibilities, we have compared the IMF-$\sigma_{e}$ trend that we have found for 
high-z dense ETGs with that of local ETGs with similar
mean stellar mass density and velocity dispersion and we have found that the IMF of massive dense ETGs does not depend on redshift. 
The similarity between the IMF-$\sigma_{e}$ trends observed 
both in dense high-z and low-z ETGs over 9 Gyr of evolution and their lower mass-normalization with respect 
to the mean value of local ETGs suggests that, independently of formation redshift, 
the physical conditions which characterized the formation of a dense spheroid \textit{on average} lead to  a
mass spectrum of newly formed stars with a higher ratio of high- to low-mass stars with 
respect to the IMF of normal local ETGs. In the direction of our findings, recent hydrodynamical simulations show that 
the higher star-formation rate that should have characterized the early stage of star formation 
of dense ETGs is expected to inhibit the formation of 
low-mass stars. Hence, compact ETGs should have higher ratio of high- to low-mass stars 
than normal spheroids, as we observe.}
{}
   
   
   
 

\titlerunning {Lower mass normalization of the stellar initial mass function for
dense massive ETGs at z $\sim$ 1.4}
   \authorrunning {Gargiulo et al.}

   \maketitle
%

\section{Introduction}

The stellar initial mass function (IMF) is the mass distribution of a stellar generation at birth; 
fixing the relative number of low- and high-mass stars formed in a burst of star formation, 
it determines the properties of the whole stellar system, 
i.e. its spectral energy distribution (SED), stellar mass, stellar mass-to-light ratio (M/L), 
chemical enrichment, and how these properties change with time.  
In fact, the most widely adopted approaches to constrain the assembly history of galaxies pass 
through the analysis of their stellar populations properties (e.g. stellar mass, age, star-formation histories) 
over the cosmic time. The properties of these stellar populations  are derived, at different redshifts, 
by comparing observations with stellar population synthesis (SPS) models which strictly depend on the IMF.
Consequently, the knowledge of the IMF is a key ingredient for our understanding of the stellar mass accretion 
over the cosmic time.
Nowadays, the local direct star counts have shown that the IMF is approximately constant throughout the disk of the Milky Way (MW) and 
can be described by two declining power laws d$N$/d$m$ $\propto$ $m$$^{-s}$, 
one  with slope s $\simeq$ 2.35 for stars with mass m $\gtrsim$ 1M$_{\odot}$,
as originally suggested by \citet{salpeter55}, and the other, 
flatter, for lower masses \citep[e.g.][]{kroupa01, chabrier03}. 
This form, measured in the MW disk, is then assumed to be universal, i.e. invariant both throughout 
the wide population of galaxies, and across the cosmic time. Nonetheless, theoretical arguments cast 
strong doubts on IMF universality \citep[see][for complete reviews]{bastian10, kroupa13}.

The Jeans mass M$_{J}$  is dependent on
the temperature T and the density $\rho$ of the gas (M$_{J}$ $\propto$ T$^{3/2}\rho^{-1/2}$)
\citep[e.g.][]{larson98, larson05, bonnell06}. The temperature of a gas is related to its metallicity Z:
at fixed heating rate, the gas cooling efficiency is reduced in absence of metals, and consequently the 
gravitational collapse of low-mass stars is inhibited in metal poor gas \citep[e.g.][]{larson05, bate05a, bate05b, bonnell06}. 
In addition to this, there is evidence that the turbulence level of the interstellar 
medium plays a major role in shaping the IMF \citep[e.g.][]{padoan02,
hennebelle08, hopkins13}.
These theoretical considerations show that the mass spectrum of the stars formed in a burst of star formation 
is deeply connected with the physical properties of the gas cloud, and hence raise doubts about the validity of the
current IMF universality assumption. 

From an observational point of view,  the IMF universality has recently been contrasted by
indirect emerging evidence in external galaxies.
The study of spectral features known to be sensitive 
to the presence of low-mass stars (e.g. the Na I doublet, the Ca II triplet) in massive local ETGs 
have shown that the strength of these gravity-sensitive features varies with ETG velocity dispersion ($\sigma$) 
and the Mg/Fe ratio in the direction of progressively bottom-heavy IMF, i.e. IMF with a lower ratio of 
high- to low-mass stars than the MW IMF, with the increase of the two parameters 
\citep[e.g.][]{conroy12,spiniello12,ferreras13,labarbera13}.  

A similar IMF-$\sigma$ trend has been found through a different (dynamical) approach based on the direct comparison of the 
\textit{true} stellar mass-to-light ratio (M/L$_{\star,true}$) with that inferred 
through, e.g. the fit of the spectral energy distribution adopting SPS models which assume
a fixed IMF as reference (M/L$_{\star,ref}$). The M/L$_{\star,true}$ has been derived by different authors 
through, e.g. detailed 2D axisymmetric dynamical models \citep{cappellari12}, 
simpler spherical dynamical models \citep{tortora13}, or through a joint lensing and dynamical 
analysis of gravitational lenses \citep{treu10}. The two independent approaches  
have been shown to achieve a similar trend with velocity dispersion \citep{tortora13, conroy13}. This
consolidates the robustness of the IMF-$\sigma$ trend in local massive ETGs, 
although caveats and limits of both methods need to be fully understood yet \citep{smith14}.

Nonetheless, the most relevant aspect for galaxy evolution, i.e. the assumption 
of a universality across the cosmic time, is still largely observationally unexplored.
Theoretically, the higher temperature of intergalactic medium, 
which is known to increase with redshift as $\propto$ 2.73(1 + z)[K], 
should contrast the formation of low-mass stars at high redshift and hence point toward a top-heavier IMF 
in early epoch with respect to that of the local Universe \citep[e.g.][]{larson98, larson05}. Similarly, 
the formation of low-mass stars should be inhibited by the absence of metals in the primordial gas 
\citep[e.g.][]{larson05}, as well as by the higher star-formation rate per unit volume observed in the early universe 
\citep[e.g.][]{larson05, dabringhausen10, papadopoulos10, papadopoulos11, narayanan12}.

Actually, in a recent analysis, \citet{shetty14} have provided the first attempt 
to constrain the IMF in massive (M$_{\star}>$10$^{11}$M$_{\odot}$) ETGs at z $\sim$ 0.8 
using axisymmetric dynamical models. They find that, on average, the IMF of z $\sim$ 0.8 ETGs is similar
to those of local ones with similar stellar mass. This result is consistent with 
the evidences that, on average, the stellar content of ETGs passively evolves from z $\sim$ 0.8 - 1 down to zero 
\citep[e.g.][]{diserego05, vanderwel05, jorgensen13, choi14}. 
In fact, if a variation of IMF in massive ETGs occurs with time, it will be mostly detectable 
through the analysis of the stellar content of ETGs at z$>$1. 
Actually, the number density of spheroidal massive galaxies increases 
by a factor $\sim$ 10 in the range 1$<z<$2 and is almost flat at lower redshift \citep{ilbert13}. 
This, coupled with the many evidences of passive evolution of the stellar content of massive ETGs in the last 7-8 Gyr, 
implies that, on average, the mean properties of the stellar content of massive ETGs observed at z $<$ 1
are not dissimilar to those observed in local ETGs. In contrast, the population of ETGs z$\sim$1.5 provides us with 
information only on those spheroidal galaxies that assembled their stellar content in the first 3-4 Gyr of the 
age of the Universe. Consequently, they retain information on the mass spectrum of stars in the early Universe.

A piece of evidence concerning the stellar content of spheroidal galaxies up to z$\sim$ 1.5, is that at fixed stellar mass M$_{\star}$,
ETGs show a remarkable spread in mean stellar mass density ($\Sigma$ = M$_{\star}$/2$\pi$ $R_{e}^{2}$).
Although conclusive results have not been achieved yet, a few recent works \citep[e.g.][]{lasker13, smith13} and 
theoretical expectations have suggested that the hotter dynamical status 
of a system could have a fundamental role in shaping the IMF \citep[e.g.][]{dabringhausen08, 
dabringhausen10, papadopoulos10, dabringhausen12, marks12}.

In this paper, taking advantage of a sample of $\sim$ 18 dense ($\Sigma > $ 2500 M$_{\odot}$pc$^{-2}$) 
ETGs we have collected at 1.26 $<z<$ 1.6 (see Section 2 for more details), 
we address the following questions: $i)$ Does dense high-z ETGs follow the same IMF-$\sigma$ trend observed for
local ETGs? $ii)$ Does the IMF of  ETGs depend on mean stellar mass density and/or on redshift? 
In Section 3 we derive the upper limit of the IMF-$\sigma$ trend of dense ETGs at $<z>$ = 1.4.
We infer the trend in dynamical masses derived via the
virial estimator.
Although this procedure is less accurate than the more sophisticated 2D axisymmetric model, in Section 
3 we discuss in detail how the source of uncertainties could affect our conclusions.
In Section 4 we present our discussion, and we summarize our results in Section 5.

Throughout the paper we adopt standard cosmology with H$_0$ = 70 kms$^-1$ Mpc$^-1$, $\Omega_m$ = 0.3
and $\Omega_\lambda$ = 0.7.

\begin{figure}[h]
\centering
	\includegraphics[angle=0,width=9.0cm]{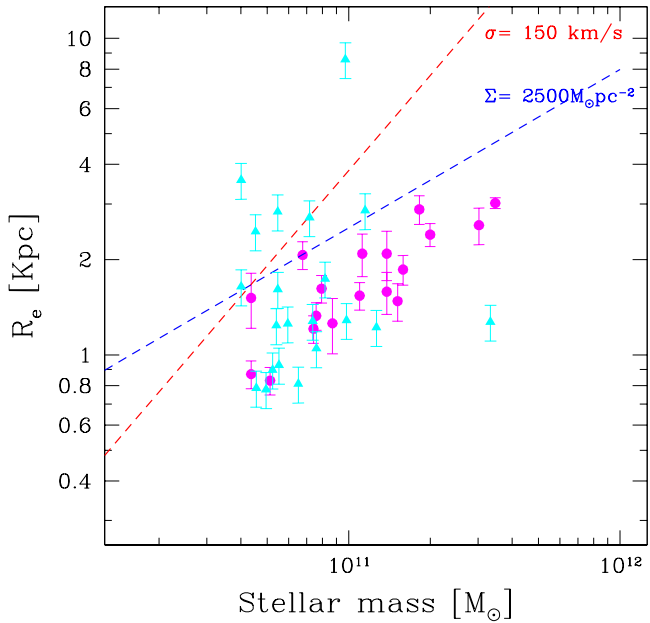} \\
	\caption{The distribution of the ETGs of the total high-z sample in the R$_{e}$-M$_{\star}$ plane (magenta points).
 Cyan triangles are a complete sample of ETGs selected from 
 the GOODS-South field in the redshift range 1.2$<z<$1.6 \citep{tamburri14} for reference.
Blue(red) lines indicate lines of constant $\Sigma$($\sigma$). We note that the sample of ETGs at $<z>$ = 1.4 
with velocity dispersions available is skewed toward more compact systems.}
	\label{sizemass}
\end{figure}

\section{The sample of dense ETGs at $<z>$ = 1.4}

The dataset we have adopted in this paper is a collection from the literature of all the 18 galaxies morphologically confirmed as ETGs at 1.26 $<z<$ 1.6, having 
stellar mass density $\Sigma > $ 2500 M$_{\odot}$pc$^{-2}$, and both 
kinematic (i.e. velocity dispersion $\sigma_{e}$) and structural (i.e. effective radius R$_{e}$ and 
stellar mass M$_{\star}$) properties available. 
For all the galaxies in the sample, the morphological classification was performed on the basis 
of a visual inspection of the galaxies carried 
out on the HST-NIC2 and WFC3 images in the F160W filter. We classified as ETGs those galaxies having a regular
shape with no sign of disk and no irregular or structured residuals resulting from the profile fitting.
We selected only those ETGs with structural parameters 
derived by fitting the HST images with a 
point spread function-convolved 2D Sersic profile and with stellar masses derived with the 
Chabrier IMF and \citet{bruzual03} models. We consider that stellar masses have a typical error of $\sim$20$\%$, 
which take into account the uncertainties related to models and photometry, but not on IMF, that for the purpose 
of our study we consider fixed. 
In Table \ref{tab1} we list all the properties of the 18 galaxies relevant 
for this analysis, i.e. the effective radii, the filter in which they were derived, 
the stellar masses, as well as their velocity dispersions
within the effective radius. We rescaled the measured velocity dispersions
to R$_{e}$ aperture following \citet{cappellari06}. 
For 14 ETGs in the sample the values of both structural and kinematical parameters
have already been published and we refer the reader to original papers for all the details (see references in Table 1), 
while for the remaining four (S2F1-511, S2F1-633, S2F1-527, S2F1-389) 
the estimates of velocity dispersions and the updated values of structural parameters 
will be presented in a forthcoming paper (Gargiulo et al. 2014 submitted). 
In the following paragraph we briefly describe the method used to derive the physical 
parameters of these four galaxies (hereafter our high-z sample galaxies).

\subsection{Our high-z sample galaxies}

We derived the velocity dispersions of the four galaxies at z$\sim$ 1.3
from spectra obtained with the ESO VLT-FORS2 spectrograph. 
The OG590 filter and the GRIS-600z grism we chose for the observations, allowed us to cover the wavelength range 
0.7$< \lambda <$ 1.0 $\mu$m with a sampling of 1.6 \AA\, per pixel. We fixed the width of the 
slit to 1'', which provides a resolution R $\simeq$ 1400, corresponding to a FWHM $\simeq$ 6.5\AA\, 
at 9000\AA. The total effective integration time per galaxy is $\sim$ 8 hours
which results in a signal-to-noise per pixel $\sim$ [8-10]. After standard reduction and 
calibration, we derived the velocity dispersions best-fitting the observed spectra with 
a set of template stars from \citet{munari05} using the software pPXF \citep{cappellari04}.
 With an extensive set of simulations we
tested the reliability of our estimates, given the wavelength region covered and the S/N of the spectra and 
found that fitting the observed spectra with a Gauss+Hermite polynomial over the 
whole range of $\lambda$, with a penalty BIAS $\simeq$ 0.1 provides reliable estimates of $\sigma$.
The effective radii were derived by \citet{longhetti07} fitting 
the HST-NIC2/F160W images with a 2D-psf convolved Sersic
profile with GALFIT \citep{peng02}. We updated the values they found  
taking in consideration the more robust estimate of the redshift based on the FORS2 spectra.
Finally, stellar masses of the four ETGs were already presented in \citet{saracco09}. However,
following the same procedure adopted by the authors, we re-estimated the stellar masses
to take into account the update value of the redshift and the now available near-IR photometry in 
the WISE RSR-W1 filter ($\lambda_{eff} \simeq$ 3.4 $\mu$m).

For 16 ETGs in the sample their effective radii were
derived fitting with a Sersic profile the WFC3/F160W-band images while for two of them structural parameters
were derived on ACS/F850LP-band images. Since high-z ETGs have significant colour gradients 
\citep[e.g.][]{gargiulo11, gargiulo12, guo11, cassata11}, in order not to affect 
the relative distribution of the effective radii of our galaxies (hence of their virial masses, see Sec 3.2),
we converted the radii of these two galaxies to those in the filter F160W
scaling the fitted values of a factor $\sim$ 0.8 \citep{cassata11, gargiulo12}.
At redshift $<z>$ = 1.4 the HST/F160W filter samples approximatively the R band restframe.
For all the galaxies in the sample, the stellar masses M$_{\star}$ were derived fitting their observed SEDs
with the SPS models by \citet{bruzual03} and the Chabrier IMF. 
Figure \ref{sizemass} shows the distribution of our sample of ETGs at $<z>$ = 1.4
in the plane defined by effective radius and stellar mass.
As a reference, we plot also a complete sample of ETGs in the redshift range 1.2 $<z<$ 1.6 \citep{tamburri14}.
The sample was extracted from the GOODS-MUSIC catalogue \citep{grazian06, santini09}  
selecting all the galaxies (1302) with K$_{s,AB} \leq $ 22. The spectroscopic redshift completeness is $\sim$ 76 per cent.
The morphological classification was performed by two authors, independently, and following the same criteria 
adopted for our spectroscopic sample. This results in a final sample of $\sim$ 196 
ETGs in the redshift range 0.6 $\leq z \leq 2.5$ from which we selected 
those at 1.26 $<z<$ 1.6 shown in Fig. \ref{sizemass}. In this redshift range the photometric 
sample is complete down to 
M$_{\star} \simeq$ 2$\times$10$^{10}$M$_{\odot}$.
 By comparing the spectroscopic sample at $<z>\sim$1.4  with the photometrically complete one, it is evident that the first 
 is skewed towards denser  systems. 
This is principally because in almost all the cases the spectroscopic targets were identified through a 
colour selection criteria which is shown to preferentially select compact ETGs  \citep[e.g.][]{saracco09,szomoru11}.
We note that, in comparison with the spectroscopic sample, the complete photometric sample  
does not include galaxies with M$_{\star} \gtrsim $ 10$^{11}$M$_{\odot}$. Actually, the number 
density of such massive ETGs at z$\sim$1.4 is extremely low ($\sim$10$^{-5}$ Mpc$^{-5}$, Ilbert et al. \citeyear{ilbert10}).
Considering the relatively small area of GOODS-South field ($\sim$140 arcmin$^{2}$), from which
the photometric sample was taken, we expect to find to $\sim$ 1.5 galaxies, as we observe. Instead, 
our high-z spectroscopic sample is a collection of spheroids usually identified as spectroscopic targets
among the most massive and red galaxies in all the deep currently available surveys. This  
increases, with respect to photometric sample, their occurrence in the range of higher stellar masses. 

\begin{table*}
\centering
\caption{Total sample of $<z>$ = 1.4 galaxies. \textit{Column 1}: ID, \textit{Column 2}: redshift, \textit{Column 3}: effective radius, 
\textit{Column 4}: sersic index, \textit{Column 5}: filter adopted to derive the surface brightness parameters, 
 \textit{Column 6}: measured velocity dispersion, \textit{Column 7}: velocity dispersion corrected to R$_{e}$ aperture following \citet{cappellari06}, 
\textit{Column 8}: Stellar mass. References for the velocity dispersion values are 
listed near the ID value. When a data of the galaxy comes from a paper different from the one where the $\sigma$ 
is published, the new reference is specified near its value.}
\begin{tabular}{cccccccc}
\hline \hline
Object              & z     & R$_{e}$             & n         & Camera-Filter  & $\sigma$           & $\sigma_{e}$         & logM$_{\star}$\\
                    &       & (kpc)               &           &                &  (km/s)            & (km/s)               & (M$_{\odot}$) \\
\hline
S2F1-511\tablefootmark{1}      & 1.267 & 2.09$\pm$0.07      & 3.3 &  HST/NIC2-F160W & 269              & 281$\pm$23            & 11.05  \\ 
S2F1-633\tablefootmark{1}      & 1.297 & 2.59$\pm$0.11      & 4.1 &  HST/NIC2-F160W & 434              & 447$\pm$27            & 11.48   \\ 
S2F1-527\tablefootmark{1}      & 1.331 & 1.57$\pm$0.24       & 3.1 &  HST/NIC2-F160W & 226              & 240$\pm$26            & 11.14      \\ 
S2F1-389\tablefootmark{1}      & 1.406 & 2.10$\pm$0.18       & 4.5 &  HST/NIC2-F160W & 224              & 234$\pm$45            & 11.14        \\ 
2239\tablefootmark{2}          & 1.415 & 2.16$\pm$0.43\tablefootmark{3} & 2.2\tablefootmark{3} & HST/ACS-F850LP & 111$\pm$35         &  116$\pm$37          & 10.64\tablefootmark{3}  \\
2470\tablefootmark{2}         & 1.415 & 1.81$\pm$0.36\tablefootmark{3} & 4.2\tablefootmark{3} & HST/ACS-F850LP & 141$\pm$26         &  150$\pm$28          & 10.94\tablefootmark{3}   \\
A17300\tablefootmark{4}        & 1.423 & 2.9$\pm$0.51\tablefootmark{5}  & 5.3       & HST/WFC3-F160W & 265$\pm$7          &  272$\pm$7           & 11.26        \\
A21129\tablefootmark{4}        & 1.583 & 1.5$\pm$0.27\tablefootmark{5}  & 5.0       & HST/WFC3-F160W & 260$\pm$9          &  278$\pm$10          & 11.18        \\
C21434\tablefootmark{4}        & 1.522 & 1.9$\pm$0.33\tablefootmark{5}  & 3.1       & HST/WFC3-F160W & 218$\pm$16         &  230$\pm$17          & 11.20         \\
C20866\tablefootmark{4}        & 1.522 & 2.4$\pm$0.42\tablefootmark{5}  & 3.0       & HST/WFC3-F160W & 272$\pm$23         &  282$\pm$23          & 11.30         \\
32915\tablefootmark{6}        & 1.261 & 1.33$\pm$0.13       & 6.3       & HST/WFC3-F160W &                    &  271$\pm$17          & 10.88        \\
2341\tablefootmark{6}         & 1.266 & 1.21$\pm$0.12       & 3.8       & HST/WFC3-F160W &                    &  196$\pm$27          & 10.87        \\
29059\tablefootmark{6}        & 1.278 & 1.62$\pm$0.16       & 4.3       & HST/WFC3-F160W &                    &  211$\pm$16          & 10.90         \\
2337\tablefootmark{6}         & 1.327 & 1.54$\pm$0.15       & 3.5       & HST/WFC3-F160W &                    &  284$\pm$20          & 11.04         \\
14758\tablefootmark{6}        & 1.331 & 0.83$\pm$0.08       & 2.2       & HST/WFC3-F160W &                    &  165$\pm$16          & 10.71        \\
5020\tablefootmark{6}         & 1.415 & 2.07$\pm$0.20       & 4.6       & HST/WFC3-F160W &                    &  180$\pm$54          & 10.83         \\
13880\tablefootmark{6}        & 1.432 & 0.87$\pm$0.08       & 2.6       & HST/WFC3-F160W &                    &  179$\pm$70          & 10.64          \\
S2F1-142\tablefootmark{7}     & 1.386 & 3.04$\pm$0.12       & 3.5       & HST/NIC2-F160W & 340$^{120}_{-60}$   & 347$^{120}_{-60}$          & 11.54        \\ 
 
\hline
 \end{tabular}
 \label{tab1}\\
\tablefoot{\tablefootmark{1}{Gargiulo et al. (2014)}, \tablefootmark{2}{\citet{cappellari09}}, \tablefootmark{3}{\citet{cimatti08}}, \tablefootmark{4}{\citet{bezanson13}},
\tablefootmark{5}{\citet{vandesande13}}, \tablefootmark{6}{\citet{belli14}}, \tablefootmark{7}{\citet{longhetti14}}, \tablefootmark{2} {\citet{vandesande13}}\\}
\end{table*}

\section{IMF-$\sigma$ trend as a function of time and mean stellar mass density}

\subsection{IMF-$\sigma$ trend in local universe}

\begin{figure*}
\centering
 \includegraphics[width=11cm]{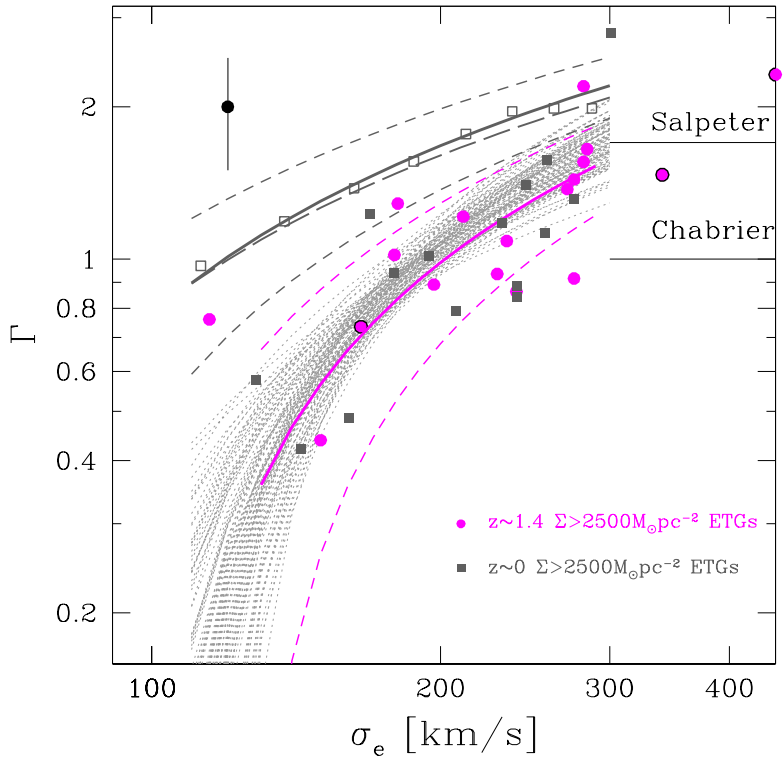}
  \caption{The trend of $\Gamma$ parameter (see Eq. 1) as a function of velocity dispersion, for normal 
(mean stellar mass density $\Sigma <$ 2500 M$_{\odot}$pc$^{-2}$) local ETGs (dark grey solid line) and 
for the whole local ETG population (long-dashed dark grey line). 
Dashed grey lines are the scatter around the best-fit relation of normal local ETGs, 
while grey open squares are the average values of the $\Gamma$s of local normal ETGs in bin of $\sim$ 50 km/s.
Magenta points are the upper limits of $\Gamma$ we have derived for high-z ETGs in our spectroscopic sample at $<z>$=1.4, 
and magenta solid line is their best-fit relation (magenta dashed lines indicate the rms around the relation).
The black small-dotted lines are the best-fit relations of 100 subsamples of local ETGs, randomly
extracted from the local sample by \citet{tortora13}, in order to have the same $\Sigma$ $and$ $\sigma_{e}$ 
distribution of our spectroscopic sample at $<z>$=1.4 (filled dark grey squares are an example of one 
local subsample). In this procedure we excluded three galaxies of the high-z sample 
(identified with magenta points with a black contour) 
since the local sample is missing ETGs with similar $\Sigma$ and $\sigma_{e}$. 
The black point indicates the typical error on $\Gamma$s of our spectroscopic sample at $<z>$=1.4. }
  \label{gammasigmaloc}
\end{figure*}

Local studies have shown that IMF in spheroidal galaxies varies with their velocity 
dispersions. 
These studies compare the stellar mass-to-light ratio M/L$_{\star,ref}$ derived through, 
e.g. the fit of the spectral energy distribution assuming a fixed reference IMF, with 
the $true$ stellar mass-to-light ratio M/L$_{\star,true}$, derived through either dynamical models or 
spectral features analysis, and investigate how their ratio $\Gamma$ = (M/L)$_{\star,true}$/(M/L)$_{\star,ref}$ 
varies with the physical properties of local ETGs.  

Usually, this IMF-$\sigma$ trend is presented in a form similar to
\begin{equation}
 \log \Gamma = \log [(M/L)_{\star,true}/(M/L)_{\star,ref}] = \alpha \log \sigma_{e} + \beta.
  \label{gamma}
\end{equation}
For the purpose of our analysis, as an example of $\Gamma$ - $\sigma_{e}$ trend in local ETGs, 
we have derived from the dataset of \citet{tortora13}
the best-fit relation of $\Gamma$ with $\sigma_{e}$ 
for normal ($\Sigma \leq$ 2500M$_{\odot}$pc$^{-2}$) local ETGs with stellar mass and velocity dispersion comparable to 
that of our high-z sample (i.e. LogM$_{\star}>$10.64 and $\sigma_{e} \gtrsim$110 km/s; 
we have already stated in the Introduction 
that all the trends found in local Universe are similar). 
Specifically, \citet{tortora13} derived, assuming different dark matter (DM) profiles, 
the $\Gamma$ values for $\sim$4500 giant ETGs
drawn from the SPIDER project \citep{labarbera10} in the redshift range of z = 0.05–0.1. 
The SPIDER sample is 95$\%$ complete at a stellar mass M$_{\star}$ = 3$\times$10$^{10}$M$_{\odot}$, 
which corresponds to $\sigma$ $\sim$ 160 km/s. In Fig. \ref{gammasigmaloc} we have reported 
the trend found assuming in the derivation of $\Gamma$ a Navarro, Frenk and White 
profile (NFW, Navarro et al. \citeyear{navarro96}) for the DM component. 
Clearly, the effect of 
reducing the DM fraction is to increase the overall stellar mass normalization 
(for a clear picture of the effect of the degree of contraction on mass normalization see 
 Dutton et al. \citeyear{dutton11}). 
In this figure, ETGs with $\Gamma$ = 1 have true stellar masses in agreement with those derived 
with the reference IMF, in this case the one by Chabrier,
while ETGs with $\Gamma >$ 1 have stellar masses greater than that 
inferred with the Chabrier IMF, i.e. their IMF normalization is more massive. 

\subsection{IMF-$\sigma$ trend for dense ETGs at $<z>$ = 1.4}

Following the same procedure adopted in the local Universe we have investigated
the IMF-$\sigma_{e}$ trend for the dense ETGs in our sample at $<z>$ =1.4. 

After defining M$_{tot}$ the total mass of the galaxy, and 
M$_{\star,ref}$ the stellar mass given in Table 1, we derived the upper limit of the $\Gamma$ parameter for 
the ETGs in our sample  as $\Gamma$ =  M$_{tot}$/M$_{\star,ref}$.
Assuming the total mass to be a proxy of the true stellar, we have implicitly assumed that these 
systems have zero DM. 

An important step in our analysis consists in deriving reliable \textit{total} masses for ETGs in our sample.
We addressed this point adopting the virial mass estimator, so for each galaxy in the sample
\begin{equation}
 M_{tot} = M_{vir} = (\beta R_{e} \sigma_{e}^{2})/G,
  \label{vir}
\end{equation}
where G is the gravitational constant and $\beta$ is equal to 
\begin{equation}
  \beta = 8.87 - 0.831 \times n + 0.0241 \times n^{2}
  \label{beta3}
\end{equation}
following \citet{cappellari06}. 

\begin{figure}
\centering
 \includegraphics[width=9cm]{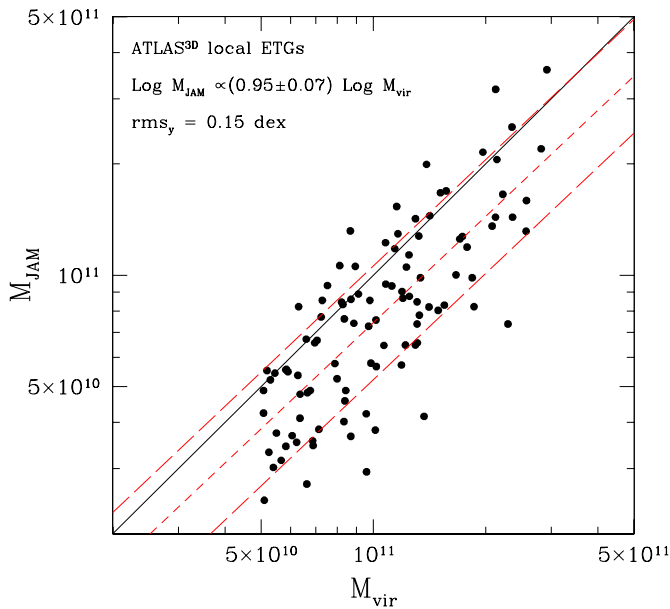}
  \caption{The dynamical mass derived with 2D axisymmetric models, M$_{JAM}$, 
versus the virial mass, M$_{vir}$, derived as in Eq. \ref{vir} for the ETGs of the local ATLAS$^{3D}$ sample.
M$_{JAM}$ was derived as stated in \citet{cappellarixv} in the caption of their Table 1. Although, on average,
Eq. \ref{vir} overestimates the dynamical masses
(black solid line is the 1:1 correlation), the offset ($\sim$ 0.2 dex) is not mass dependent.}
  \label{estvir}
\end{figure}

Typical errors on our $\Gamma$ estimates are $\sim$0.5. 
Detailed estimates of dynamical masses based on more sophisticated 2D axisymmetric dynamical models 
of local ETGs have shown that the virial Eq. \ref{vir} could overestimate the total masses \citep[][]{cappellarixv}.
In addition to the uncertainties on the $true$ value of virial mass for each ETG, to which we will return later, 
we point out that 
the $relative$ values of M$_{vir}$ are not expected to change. 
In Figure \ref{estvir}, the 
total dynamical mass of local ATLAS$^{3D}$ ETGs \citep{cappellarii} derived with accurate 2D axisymmetric models (M$_{JAM}$) is 
given versus their virial mass M$_{vir}$ derived as in Eq. \ref{vir}.
The two mass estimates were derived from the ATLAS$^{3D}$ database as follows. The M$_{JAM}$ masses, following 
\citet{cappellarixv} (see caption of their Table 1), were derived as (M/L)$_{JAM}\times$ L, where (M/L)$_{JAM}$ 
and L are the mass-to-light ratio obtained
via dynamical models and the galaxy total luminosity, found in Cols. 6 and 15 of their Table 1, respectively. 
For the estimate of M$_{vir}$ for ATLAS$^{3D}$ ETGs, we adopted for R$_{e}$ and $n$ 
those derived from the 1D Sersic fit of the observed light profiles of the galaxy
(R$_{e,1D}$ and  $n_{1D}$, Cols. 3 and 4 of the Table C.1 in Krajnovi\'{c} et al. \citeyear{cappellarixvii}). 
Although the procedure they adopted to derive the R$_{e}$ is different from the 2D approach of GALFIT, 
the authors show that, at least for the Sersic index, the two procedures return similar results, with a difference of just 0.08 
between the median values of the two distributions. 
For what concerns the $\sigma$, we referred to the stellar velocity dispersion 
measured by co-adding all SAURON spectra contained within the physical region of 1kpc (Col. 4 of 
Table 1 in \citet{cappellarixv}). We corrected this measure to those within R$_{e,1D}$ following 
\citet{cappellari06}. 
In Figure \ref{estvir} we plot the ATLAS$^{3D}$ ETGs with 10.7 $\lesssim Log M_{vir} \lesssim$ 11.5 in order 
to approximately match the mass range of our spectroscopic sample. 
In this mass range, on average, we observe that Eq. \ref{vir} overestimates the dynamical masses, 
but the offset ($\sim$ 0.2 dex) is not mass dependent. 
The values of virial mass estimated via Eq. \ref{vir}, therefore, are \textit{relatively} 
reliable and hence, so are the $\Gamma$ parameters 
of our high-z ETGs.
Based on these considerations, and looking at the distribution of high-z dense ETGs in the $\Gamma$-$\sigma_{e}$ plane,
the first result we have found is that the $\Gamma$ of high redshift dense ETGs show a $trend$ with 
their $\sigma_{e}$ (Spearman's rank coefficient 0.84, probability 6.9$\times$10$^{-5}$)
and that this trend is very similar to that observed in the
local Universe, that is, their IMF becomes more massive with increasing velocity dispersion. 
In particular, in the range of velocity dispersion $\sim$ [120-300] km/s we have found 
$\Gamma$ = 2.99 Log$\sigma_{e}$ - 5.22  with a scatter of 0.28.

The possible uncertainties on the true value of M$_{vir}$ do not affect this result, since 
the $trend$ is dependent on relative values. Thus, within the approximation of our analysis, the trend of IMF 
with $\sigma_{e}$ is also observed in high-z ETGs.

The second aspect we want to investigate is whether the IMF of ETGs depends on redshift and/or on mean stellar mass density.
To address this second question we have compared the \textit{normalization} of the IMF-$\sigma_{e}$ trend we have found 
for dense ETGs at z $\sim$1.4  with that observed in local universe.

Although we are making the extreme assumption of no dark matter, and thus the $\Gamma$ we have derived 
for high-z ETGs are upper limits, the normalization of the  IMF-$\sigma_{e}$ trend for dense high-z ETGs 
is at 1$\sigma$ lower than that observed for normal local ETGs. In particular, on average,
at fixed velocity dispersion, the $\Gamma$s of high-z dense ETGs are a factor of $\sim$ 2
lower than the average value observed for normal local ETGs\footnote{Since the $\Gamma$ of a given galaxy increases 
for decreasing amount of dark matter in the model, it could be more intuitive to expect
that the IMF-$\sigma_{e}$ trend we have derived had a higher mass-normalization than that of typical local ETGs.
However, by definition, the $\Gamma$ of a galaxy is  $\Gamma$ = M$_{true}$/M$_{ref}$ = $\frac{R_{e}\beta(1- f_{DM})\sigma_{e}^{2}}{G M_{ref}}$
where f$_{DM}$ is the galaxy dark matter fraction.
Thus, if we consider two different galaxies with the same velocity 
dispersion $\sigma_{e}$, the values of their $\Gamma$s are not simply 
related to their dark matter fractions, but also to their M$_{ref}$/R$_{e}$ ratio, 
hence to their mean stellar mass density. Thus, the position occupied by a galaxy in the IMF-$\sigma$ plane
does not only depend on its f$_{DM}$, but also on its R$_{e}$/M$_{\star}$ ratio.}.
This could be ascribed to an evolution of IMF with redshift or to a dependence of IMF on mean stellar mass density.
To discriminate between the two possibilities, we have compared the IMF-$\sigma_{e}$ trend of high-z dense ETGs with that
of equally dense local ETGs.
In fact, if the IMF of massive dense ETGs depends on redshift, 
we should  note that the average IMF-$\sigma_{e}$ trend of local dense ETGs differ from the one we observe at high-z. 
Specifically, to address the topic, we randomly extracted from the sample of \citet{tortora13}, 
100 subsamples of local ETGs with Log[M$_{\star}/M_{\odot}]>$10.64 (as the high-z sample) and the same distribution in 
$\Sigma$ and $\sigma_{e}$ of our spectroscopic high-z sample. We checked the goodness of the local subsamples
in reproducing the high-z sample with a KS test. In the random extraction, we retained a local subsample 
if the probability that its distribution of $\Sigma$ $and$ $\sigma_{e}$ and that of the high-z sample are 
extracted from the same parent distribution is higher that 60$\%$. 
In this procedure we excluded three galaxies of the high-z sample 
since the local sample is missing ETGs with similar $\Sigma$ and $\sigma_{e}$. 
For each of these 100 subsamples we derived the
best fitting $\Gamma$-$\sigma$ relation in the range $\sim$ [120-300] km/s.
The resulting relations are shown in Fig. \ref{gammasigmaloc}.
Actually, the comparison shows that the IMF trend of dense high-z ETGs is consistent with that of local ETGs with similar 
velocity dispersion and mean stellar mass density.

However, we notice that our estimates of $\Gamma$ are based on two assumptions, namely 
that on the DM and on the Eq. \ref{vir} as a reliable estimator of total mass. Although these assumptions
do not affect the relative values of total masses (as we have shown above), they could have a significant 
impact on their \textit{absolute} values, and hence on the normalization of the IMF-$\sigma_{e}$ trend we observe.
With regard to our assumption of zero dark matter fraction, 
we note that dense ETGs are expected to be stellar matter dominated 
\citep[e.g.][]{conroy13}, and, on average, the dark matter fraction within the effective radius 
in massive dense ETGs is found to be $<$10$\%$ (Tortora, private communication). 
Similar results are derived also from ATLAS$^{3D}$ survey. Specifically, we checked the dark matter 
fraction f$_{DM}$ of ATLAS$^{3D}$ ETGs (taken form \citet{cappellarixx}) with 
10.7 $\lesssim Log M_{vir} \lesssim$ 11.5 and $\Sigma >$ 2500 M$_{\odot}$pc$^{-2}$ (for a detailed
description of $\Sigma$ estimates for ATLAS$^{3D}$ ETGs see below) and found that f$_{DM}$ = 0.11 $\pm$ 0.11.
Moreover, simulations of dissipationless mergers of spheroidal galaxies 
have shown that this fraction can only decrease going back with time, 
pointing toward an even lower fraction of dark matter in high-z spheroidal galaxies \citep{hilz13}.

However, the values of $\Gamma$ we have found for high-z systems could be affected by our uncertainties 
on true values of M$_{tot}$, hence of M$_{vir}$. In the following using the results obtained through 2D dynamical models of local ATLAS$^{3D}$ ETGs 
we show that Eq. \ref{vir} provides reliable absolute estimates of total masses for dense ETGs at z$\sim$1.4.
\citet{cappellarixv} in their Figure 13 and 14, by comparing the estimates of total mass of local ETGs derived 
through accurate 2D dynamical models (M$_{JAM}$) with those inferred from Eq. \ref{vir}, 
clearly showed that the reliability of total masses estimated using the virial theorem is closely 
related to the choice of virial coefficient $\beta$ 
and to the accuracy of R$_{e}$ estimates, which depends on the quality of the data and on the 
technique adopted to derive them.
Following a similar approach, analogous results have been found by \citet{shetty14} for massive galaxies at z$\sim$1.
As far as the robustness of size measurements of high-z ETGs, many works \citep[e.g.][]{szomoru12, davari14}
have addressed the topic, showing that the typical approach adopted at high-z, i.e. 2D Sersic-fit
of the light profile out to infinity on deep space-based images,
is able to provide accurate R$_{e}$ estimates (uncertainties of $\sim$15$\%$).
In the case in which radii are derived through this approach (i.e. through a Sersic-fit of the light profile out to 
infinity), \citet{cappellarixv} showed that the best approximation of dynamical masses is recovered assuming 
in the virial equation the non-homologous $\beta$ coefficient of Eq. \ref{beta3} 
rather than a constant virial coefficient (see the two right panels in their Fig. 14). 
Actually, in these conditions,  
M$_{vir}$ shows systematic offset of $\sim$ 0.2 dex with M$_{JAM}$ with a scatter 0.15 dex
as we have shown in Fig. \ref{estvir}.
To further exploit the origin of this offset we compared for ATLAS$^{3D}$ ETGs the value of their $\beta$
coefficient, derived following Eq. \ref{beta3}, with those inferred from their 2D-dynamical models (K) as
\begin{equation}
 K = (M_{JAM} * G) / (R_{e}\sigma_{e}^{2}),
\label{kappa}
\end{equation}
where M$_{JAM}$ was derived as described above, R$_{e}$ is R$_{e,1D}$ (Table C.1 from \citet{cappellarixvii}) 
and $\sigma_{e}$ is the velocity dispersion within 1 kpc (Table 1 \citet{cappellarixv}) corrected
to the R$_{e,1D}$ aperture.

\begin{figure}
\centering
 \includegraphics[width=9cm]{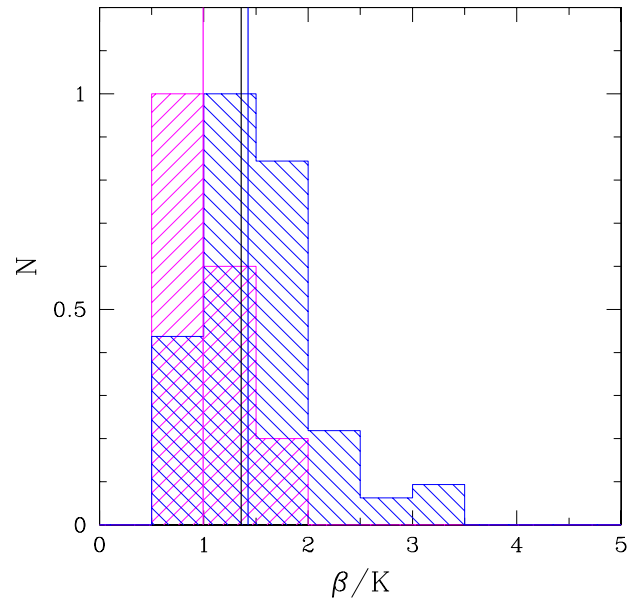}
  \caption{The distribution of the ratio $\beta$/K for ATLAS$^{3D}$ ETGs with 
$\Sigma \geq$ 2500M$_{\odot}$pc$^{2}$ (magenta histogram) and for those with $\Sigma <$ 2500M$_{\odot}$pc$^{2}$ 
(blue histogram). Coloured solid lines are the median
values of the two relative distributions, while black one refers to the total sample. We have included in the plot only local ETGs with stellar masses in the 
same range of dense high-z ETGs in our sample 
(10.7 $\lesssim$ LogM$_{vir}$/M$_{\odot}$ $\lesssim$ 11.5). The $\beta$ parameter was derived through Eq. \ref{beta3}, while the K coefficient 
through Eq. \ref{kappa}.}
  \label{beta}
\end{figure}

In Fig. \ref{beta} we plot the distribution of the ratio $\beta$/K for ATLAS$^{3D}$ ETGs with 
$\Sigma \geq$ 2500M$_{\odot}$pc$^{2}$ and for those with $\Sigma <$ 2500M$_{\odot}$pc$^{2}$. 
To estimate the mean stellar mass density of ATLAS$^{3D}$ ETGs we adopted for R$_{e}$ the R$_{e,1D}$ values 
and derived their stellar masses multiplying the stellar mass-to-light ratio (Table 1 from \citet{cappellarixv}) 
by the total luminosity L (Table 1 from \citet{cappellarixv}). 
Since the ATLAS$^{3D}$ stellar mass-to-light ratio refers to the Salpeter IMF,
we converted the stellar masses to the Chabrier IMF following \citet{longhetti09}.

We have included in the plot of Fig. \ref{beta} only local ETGs with 
10.7 $\lesssim$ LogM$_{vir}$/M$_{\odot}$ $\lesssim$ 11.5.
Although, on average, for the total sample the $\beta$ coefficient overestimates 
the virial masses of a factor $\sim$ 1.35, this  is principally due to normal  ETGs
(i.e. ETGs with $\Sigma <$ 2500M$_{\odot}$pc$^{2}$). In fact, a clear trend
of the $\beta$ parameter with the compactness of the system emerges: while the majority of normal ETGs shows a 
$\beta$/K ratio $>$ 1 (median value 1.42$^{-0.36}_{+0.37}$, where errors indicates the 25th and 75th percentile), 
for dense ETGs ($\sim$ 20 ETGs on a total of 103 ETGs in the figure) the virial masses estimated by means of Eq. \ref{vir} 
are perfectly in agreement with those inferred through more sophisticated dynamical models 
(median value of $\beta$/K = 0.99$^{-0.11}_{+0.09}$).
This result could be expected given the emerging picture on the variation of ETGs structure in 
the R$_{e}$ - M$_{\star}$ plane. Actually, \citet{cappellarixx} in their schematic summary of Fig. 14 clearly showed
that in the mass range 3$\times$10$^{10} <$ M$_{\star} <$ 2$\times$10$^{11}$ ETGs are dominated by the fast rotators, 
with very few slow rotators. However, at fixed stellar mass, the relevance of the bulge steadily increases with the mean stellar mass density.
We argue, therefore, that 
the assumption of spherical isotropic models in the derivation of the $\beta$ coefficient could fit better 
for dense systems than for normal ETGs, as can be seen in Fig. \ref{beta}.
Clearly, this result could not be necessarily valid for high-z dense 
ETGs which could be structurally and dynamically different from local ones. However, 
it has been found that both the kinematical and structural properties of dense high-z ETGs are very similar 
to those of equally dense local ETGs, and their $\beta$ parameters should be similar as well. \citep[e.g.][]{saracco14, trujillo13, ferremateu12, vanderwel12, buitrago13}.
In this context, very interestingly, the mean value of the $\beta$ parameter we have derived for our dense 
high-z ETGs is 7.0 $\pm$ 1.7, and
perfectly agrees with the mean value of massive ATLAS$^{3D}$ dense ETGs, 7.5 $\pm$ 1.7. 

To conclude, we have found that dense high-z ETGs $i)$  follow the same IMF-$\sigma$ trend of local ETGs, 
$ii)$ have a lower mass normalization with respect to the average IMF-$\sigma$ trend of normal local ETGs population, and 
$iii)$ their mass-normalization is consistent with that of local ETGs with similar 
velocity dispersion and mean stellar mass density. 
In fact, neither our assumption about the dark matter fraction or the estimate of dynamical masses 
through virial estimator affect this result. Our finding of a lower mass normalization for dense high-z ETGs is 
in agreement with the recent result obtained through lens modeling of a z =
1.62 compact (R$_{e}$ $\simeq$ 1.3kpc, and M$_{\star}$ $\simeq$ 2$\times$10$^{11}$M$_{\odot}$) 
early-type gravitational lens galaxy \citep{wong14}. Although the constraints are broad, 
the authors have found that the stellar mass of this galaxy is more consistent
with that derived with a Chabrier IMF than with a Salpeter IMF.

In term of stellar content, these findings imply that at fixed velocity dispersion and at any redshift,
the IMF of dense ETGs has a higher ratio of high- to low-mass stars, with respect to the IMF of typical local ETGs.
In particular, assuming that the IMF is modelled as a single power law
d$N$/d$m$ $\propto$ $m$$^{-s}$ in the range of stellar masses [0.1-100]M$_{\odot}$, we found that its slope
varies from s = $\sim$1.5 for high-z dense ETGs with velocity dispersion $\sim$150 km/s to s = 2.35 (hence the 
Salpeter) for ETGs with velocity dispersion $\sim$ 300 km/s.

\section{The IMF in massive spheroidal galaxies: Does it depend on $when$ or on $how$ ETGs assembled their stellar mass?}

The similarity between the IMF trend of high-z and low-z dense ETGs over 9 Gyr of evolution and their lower 
normalization with respect to that of a typical local ETG suggest that i) the IMF in massive dense ETGs does not depend 
on the \textit{epoch} at which they accreted their stellar mass and that ii) the physical conditions responsible for 
the higher mean stellar mass density of dense ETGs, at any redshift, favour a mass spectrum of new stars
 with a higher ratio of high- to low-mass stars with respect to that of a normal local ETGs. 
 
Our current knowledge of dense ETGs shows that the compactness of a spheroidal galaxy is closely
related to its star-formation rate (SFR). Local studies have shown that the FP residuals of local ETGs 
are anticorrelated both with galaxy age 
\citep[e.g.][]{forbes98, reda05, gargiulo09, graves09} and with the $\alpha$-element abundance ratio, 
$\alpha$/Fe \citep[e.g.][]{gargiulo09, graves09} in trends whereby ETGs 
more compact than expected from the FP relation have stellar populations systematically older 
and with higher abundances than average. However, 
a multiple regression analysis of FP residuals showed 
that $\alpha$/Fe is the driving parameter, since there is $no$ age correlation with compactness at fixed $\alpha$/Fe, but
a very strong $\alpha$/Fe correlation at fixed age \citep[e.g.][]{gargiulo09}.
It has been long realized that the stellar $\alpha$/Fe ratio is a 
robust indicator of the timescale $\tau$ of the star formation \citep[e.g.][]{tinsley80, greggio83, matteucci96}, 
with higher abundance for shorter timescale. Thus, 
local results show that the compactness of a spheroidal galaxy is not the imprint
of $when$ it had assembled its stellar mass, but mostly of $how$, and
more specifically, of the $duration$ of its mass assembly history. 

These findings show that the higher the SFR at which a galaxy assembles its stellar mass, the higher
its compactness. Recent hydrodynamical simulations have shown that, in a galaxy, 
star formation at a rate greater than few M$_{\odot}$yr$^{-1}$ leads to an increase of the Jeans mass of the system
\citep{narayanan12}, and thus inhibits the formation of low-mass stars.
This result agrees with our findings. With respect to normal ETGs,
the higher SFR that caracherizes the formation of dense ETGs, independently of z, 
should lead to a lack of low-mass stars with the consequent increase in the ratio of high- to low-mass stars, 
as we observe.

To summarize, on the basis of our results we have shown that high-z dense ETGs follow 
the same IMF-$\sigma$ trend of local dense ETGs. This implies that 
the IMF of massive dense ETGs do not depend on time. 
Independently of z, the observed lower mass-normalization is a characteristic of ETGs with 
higher mean stellar mass density. 
If this result is extendible to all ETGs, that is, if
the mass normalization is linked to the mean stellar mass density of the galaxies, then, at any  redshift, 
the IMF of massive ETGs will have a higher or lower ratio of high- to low-mass stars according 
to the peculiar physical conditions in which a galaxy assembled its stellar mass.

Starting from these results, coupled with the recent findings on normal and dense ETGs, in the 
following we discuss the implication that our findings can have on the evolution with time of the $mean$ IMF 
of ETGs population.
Recent evidence suggests that the first spheroids formed in the early Universe, were dense
\citep[e.g.][]{saracco10,cassata13,vandokkum10}.
At later epochs, compact ETGs
continued to appear, together with normal ETGs, at least down to z $\sim$ 1. 
In fact, the first studies of the number density of dense and normal massive quiescent galaxies have shown 
that 
dense spheroids were numerically dominant at z$>$1 while normal ETGs are
more common at z$<$1 \citep{cassata13}. 
If the mass normalization of IMF is related to the mean stellar mass density, 
with densest systems those with a lower mass-normalization as we 
have found, the evolution of number density of dense and normal ETGs should imply a 
trend of the $average$ IMF of massive ETGs population with z, in the direction of a higher mass-normalization with 
decreasing redshift. On this basis, consistent with the many theoretical predictions on the evolution of Jeans mass with 
redshift (see Introduction), our result implicitly suggests that the mean IMF of ETGs population in 
the early universe had a higher abundance of high-mass stars than the mean 
IMF of local ETG (assuming a power-law IMF).
Clearly, this proposed picture needs to be further constrained by observations aimed at investigating 
the IMF in complete sample of massive ETGs at different epochs.

\section{Summary and conclusions}

We have used a spectroscopic sample of dense
($\Sigma > $2500M$_{\odot}$pc$^{-2}$) ETGs we collected at z=1.4
to investigate the variation of the IMF of massive ETGs
as a function both of $\Sigma$ and z.
The results of our analysis are the following:
\begin{itemize}
 \item Massive dense ETGs at z=1.4 follow the
same IMF-$\sigma_{e}$ trend of typical local
ETGs, but with a lower mass-normalization.
If massive dense ETGs were the first spheroids to form, as recent findings suggest, this result shows that 
the \textit{average} normalization of the IMF
of the first massive ETGs  is lower than that of a Chabrier IMF. 
This lower normalization could be an evidence of an evolution
of the IMF with time or of a correlation with $\Sigma$
\item We find that the IMF-$\sigma_{e}$ trend of dense ETGs at z=1.4
is consistent with those of local ETGs with similar velocity
dispersion and mean stellar mass density.
This result is an evidence that the IMF, at least for dense ETGs,
is not time dependent. 
\end{itemize}
Irrespective of the formation redshift, at fixed velocity
dispersion, the physical conditions which characterized
the formation of dense ETGs lead to a mass spectrum of
newly formed stars with a higher ratio of high- to low-mass stars
than the IMF of typical ETGs of similar velocity
dispersion.
In facts, hydrodynamical simulations have shown that the Jeans
Mass increases with the star-formation rate (SFR). Thus,
given the higher SFR that should characterize
the early phases of star formation of dense ETGs, we suggest that the lower mass-normalization
we observe in dense ETGs is the result of a lack of low-mass stars.

On this basis, we suggest that the choice of the proper
IMF for an individual massive ETG is not dependent on the time
at which it assembled its stellar mass, but mostly 
on the efficiency of its star formation, hence on $\Sigma$.

\begin{acknowledgements}
The authors warmly thank the referee for his/her helpful comments and suggestions, 
and C. Tortora for the scientific discussions and for providing us with local data.
This work has received financial support from Prin-INAF
1.05.09.01.05 and is based on observations made with ESO Very Large Telescope 
under programme ID 085.A-0135A.
\end{acknowledgements}

\nocite{}
\bibliographystyle{aa}
\bibliography{paper_gargiulo}

\end{document}